# Mind the Gap: Missing Cyber Threat Coverage in NIDS Datasets for the Energy Sector

Adrita Rahman Tory[1], Khondokar Fida Hasan[2*], Md Saifur Rahman[1], Nickolaos Koroniotis[2], and Mohammad Ali Moni[3]

[1] Bangladesh University of Business and Technology (BUBT), Mirpur-2, Dhaka-1216, Bangladesh
[2] University of New South Wales (UNSW), ACT 2601, Australia
[3] University of Queensland (UQ), St Lucia, Brisbane, QLD 4072, Australia

**Abstract.** Network Intrusion Detection Systems (NIDS) developed using publicly available datasets predominantly focus on enterprise environments, raising concerns about their effectiveness for converged Information Technology (IT) and Operational Technology (OT) in energy infrastructures. This study evaluates the representativeness of five widely used datasets: CIC-IDS2017, SWaT, WADI, Sherlock, and CIC-Modbus2023 against network-detectable MITRE ATT&CK techniques extracted from documented energy sector incidents. Using a structured five-step analytical approach, this article successfully developed and performed a gap analysis that identified 94 network observable techniques from an initial pool of 274 ATT&CK techniques. Sherlock dataset exhibited the highest mean coverage (0.56), followed closely by CIC-IDS2017 (0.55), while SWaT and WADI recorded the lowest scores (0.38). Combining CIC-IDS2017, Sherlock, and CIC-Modbus2023 achieved an aggregate coverage of 92%, highlighting their complementary strengths. The analysis identifies critical gaps, particularly in lateral movement and industrial protocol manipulation, providing a clear pathway for dataset enhancement and more robust NIDS evaluation in hybrid IT/OT energy environments.

## 1 Introduction

Cyber operations targeting critical infrastructure, particularly in the energy sector, have increased significantly in both frequency and complexity. Prominent incidents such as Stuxnet's manipulation of uranium enrichment controllers [9], Triton's disablement of safety instrumented systems in a Saudi petrochemical plant [14], and repeated Sandworm attacks disrupting the Ukrainian power grid, exemplify the potential of malware to cause extensive physical damage. This escalation is exacerbated by the convergence of Information Technology (IT) and Operations Technology (OT), characterized by legacy equipment, extended lifecycles, stringent safety regulations, and increasing digital connectivity, collectively amplifying the cyberattack surface [19].

---

* Corresponding author, *Email: fida.hasan@unsw.edu.au*

Network based intrusion detection and prevention systems (IDS/IPS) are essential defensive mechanisms deployed in industrial environments to mitigate such cyber threats. Guidelines, such as those of NIST, emphasize the importance of continuously monitoring the traffic of the control system to detect unauthorized activities and policy violations [19]. Despite recent advances that leverage artificial intelligence techniques to enhance IDS performance, existing solutions often lack robust training data. Specifically, these datasets often omit detailed representations of industrial protocols, such as Modbus, IEC 61850, or DNP3, which are crucial to detect protocol-level anomalies accurately. In addition, they fail to adequately capture operational semantics, which refers to the contextual understanding of normal and abnormal industrial processes, which is essential for identifying subtle, yet critical deviations. Furthermore, the complexity of cross-domain attack paths, involving lateral movements from enterprise IT networks into sensitive OT environments, remains inadequately represented, significantly limiting the effectiveness of real-time threat detection [17,3,12,18,2].

To address these shortcomings, this article introduces a systematic and quantitative framework to assess the coverage gaps of publicly available IDS datasets. Our proposed framework specifically evaluates five widely cited datasets: CIC-IDS2017, SWaT, WADI, Sherlock, and CIC-Modbus2023, against 94 network-detectable attack techniques identified from 274 documented energy sector incidents within the MITRE ATT&CK knowledge base. This assessment involves prioritizing techniques based on incident frequency and practitioner-developed risk indices [11], filtering for network observability, profiling each dataset's protocol and scenario coverage, and validating technique-to-dataset mappings using advanced large-language models. By bridging gaps between enterprise and industrial environments, this approach addresses the fragmentation and blind spots in current IDS datasets and enhances the ability of detection tools to model and respond to complex, multistage cyber threats.

The primary contribution of this paper is to propose a structured methodology for systematically identifying and quantifying dataset coverage gaps relevant to hybrid IT/OT environments. Subsequently, it also presents a detailed mapping and analysis, revealing significant deficiencies in current datasets, especially concerning lateral movement and industrial protocol manipulation, and provides a prioritized set of scenarios for dataset augmentation. These contributions provide a practical approach to enhance the realism and effectiveness of IDS evaluations in the cybersecurity of the energy sector.

The remainder of this paper is organized as follows. Section 2 reviews the background and related work. Section 3 describes the proposed methodology. Section 4 presents the experimental evaluation through a case-study application. Section 5 discusses the results, and Section 6 concludes the paper with key findings and future directions.

## 2 Background Study

### 2.1 Intrusion-detection requirements in modern energy infrastructure

Industrial Control Systems (ICS) in energy networks, originally prioritizing safety and availability, face escalating cyber threats due to increasing digitalization and integration of SCADA systems with IT services and cloud analytics. This convergence eliminates traditional air-gaps, expanding the attack surface. NIST SP 800-82 Revision 3 notes that the long operational lifecycles of energy equipment and the drive for remote monitoring create significant vulnerabilities for adversaries [19]. Malicious cyber capabilities now enable direct physical disruption, moving beyond mere reconnaissance.

Cybersecurity standards, including NIST SP 800-82 Rev. 3 [19] and IEC 62443-3-3 [6], mandate multi-layer, network-based intrusion detection for critical energy operations. These frameworks require operators to acquire network traffic via TAPs and SPAN ports and implement continuous monitoring across enterprise, control center, and field layers to detect lateral movement before physical impact. This regulatory consensus dictates NIDS sensor placement.

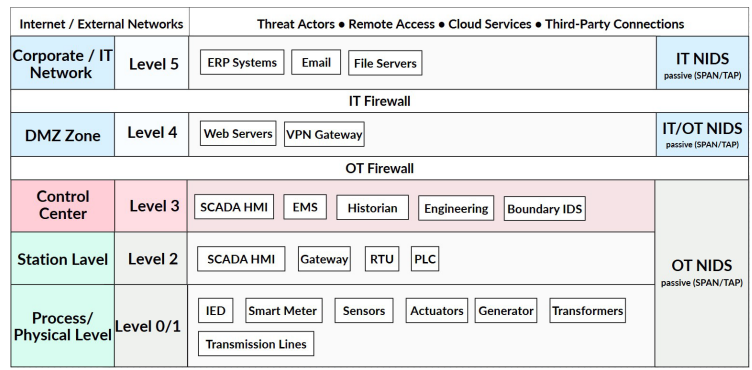

**Fig. 1.** Layered Energy Infrastructure Architecture with NIDS Deployment Zones

Empirical attack chains validate this layered approach. The 2015 Black-Energy compromise of Ukrainian distribution utilities involved initial spear-phishing in the corporate zone, pivoting through the DMZ, and ultimately issuing malicious breaker commands within substations [7]. The 2016 CrashOverride/Industroyer incident demonstrated protocol-level sabotage of IEC 101/104 messages once adversaries accessed Level 2 networks [8]. Sensors at the enterprise boundary detect credential abuse, DMZ devices observe command-and-control beacons, and process-bus monitors capture illicit field-bus traffic, all passively via TAPs as per NIST mandates [19]. These placements align with IEC 62443's zones and conduits model, addressing past vulnerability exploits.

Detection objectives vary by layer: Level 4 monitors focus on phishing payloads and malware signatures; Level 3 sensors flag suspicious SMB or RDP pivots; and Level 1/0 devices inspect industrial protocol manipulation, such as forged GOOSE trip commands. Process-level inspection has stringent timing constraints; IEC 61850-5 caps end-to-end GOOSE latency for protection traffic at 4 ms [19]. Utility design guides confirm that gigabit-Ethernet process buses meet this, with Cisco's architecture supporting 10 Mb to 1 Gb links for GOOSE while ensuring less than 4 ms delivery [6]. Therefore, NIDS engines for Levels 1/0 must handle ≥1 Gb s$^{-1}$ throughput and perform deep parsing of IEC 61850, DNP3, and Modbus frames within latency limits. To ensure model efficacy, public datasets supporting process-layer detection must replicate these bandwidth and timing properties.

The layered mandates, empirical threat paths, and real-time constraints establish a clear benchmark for evaluating dataset suitability. Therefore, this study assesses publicly available NIDS datasets based on zone coverage, protocol mix, and traffic-rate realism, aligning dataset characteristics with operational needs to determine their practical readiness for defending modern energy infrastructures.

### 2.2 IDS Training Datasets

State-of-the-art network intrusion detection systems increasingly rely on anomaly-based machine learning models that flag statistical deviations from a learned baseline rather than matching static signatures [17]. Convolutional neural network detectors, for example, achieve low false positive rates in industrial datasets when trained in suitable labeled traffic [16]. These models cannot learn meaningful decision boundaries without a large, balanced dataset that reflects their deployment environment's protocol mix, traffic rates, and attack tactics [20]. Consequently, weaknesses in the underlying dataset—such as class imbalance, obsolete protocols, or simulated artifacts—directly erode model reliability and elevate operational risk [13].

Empirical studies of benchmark suites such as UNSW-NB15 demonstrate that training on traffic that omits modern low-footprint attacks inflates apparent detector accuracy while masking real-world blind spots [17]. Similar concerns have been raised for CIC-IDS2017, where protocol coverage remains largely enterprise-centric despite meticulous labeling [4].

To evaluate whether existing resources satisfy the layered requirements outlined earlier, Table 1 juxtaposes the most-cited public datasets against four criteria: domain scope, industrial-protocol support, attack breadth, and known limitations.

The first four entries in Table 1 originate from enterprise-network research and contain no industrial payloads. Although UNSW-NB15 extends KDD Cup 99 with modern exploits, both datasets remain blind to field-bus manipulation, making them ill-suited for Levels 1/0 monitoring in energy infrastructures [20,16]. CIC-IDS2017 and its 2018 successor improve traffic realism and labeling quality yet still focus on HTTPS, SSH, and related enterprise protocols, leaving critical IEC or DNP3 semantics unrepresented [4].

SWaT and WADI focus on industrial control by providing packet captures from water-treatment and distribution plants. Both include labeled Modbus exchanges and dozens of process-level attacks, offering valuable ground truth for anomaly research [10,1]. Nevertheless, their single-site scope and absence of enterprise-pivot traffic mean models trained exclusively on these sets risk false negatives when adversaries traverse corporate zones before manipulating pumps or valves.

**Table 1.** Comparison of Public IDS Datasets by Domain, Protocols, and Coverage

| Dataset | Year | Domain | Industrial Protocols | Principal Attack Classes | Key Limitations | Ref. |
|---|---|---|---|---|---|---|
| UNSW-NB15 | 2015 | IT | None | 9 contemporary IT attacks | No OT payloads | [16] |
| CIC-IDS2017 | 2017 | IT | None | Brute-force, DoS/DDoS, Botnet, Web, Heartbleed, Infiltration | Flow level only; IT-centric | [4] |
| SWaT | 2016 | OT | EtherNet/IP (CIP) | 36 cyber-physical attacks | Single water-treatment plant | [10] |
| WADI | 2017 | OT | Modbus/TCP | 14 distribution-network attacks | Small-scale testbed | [1] |
| CIC-Modbus 2023 | 2023 | OT | Modbus/TCP | 9 substation Modbus attacks | No enterprise pivot traffic | [5] |
| Sherlock | 2025 | Hybrid (power grid) | IEC 60870-5-104 | Multi-stage grid attacks | Simulated; early release | [21] |

Recent initiatives target the power-system context. CIC-Modbus2023 introduces simulated substation traffic with nine Modbus-specific techniques but still omits higher-layer IEC protocols and business-IT chatter [5]. The Sherlock dataset combines IEC 60870-5-104 exchanges with process telemetry across three grid scenarios. Yet, its reliance on a co-simulator means physical-layer noise and operator workloads remain approximations [21].

These observations reveal a persistent mismatch between publicly available datasets and the layered, time-critical traffic observed in modern energy infrastructure.

## 3 Methodology

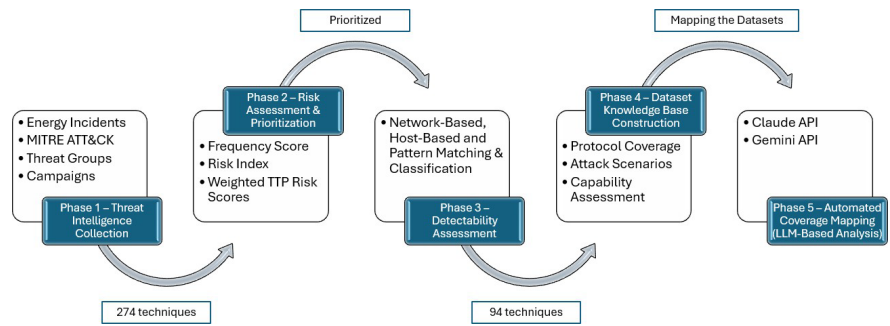

**Fig. 2.** Systematic Workflow for Energy Sector Attack Technique Coverage Assessment

The study converts incident-driven threat intelligence into a quantitative gap-analysis workflow (Figure 2) that balances empirically derived risk with the practical observability limits of network-based IDS datasets. The five phases and their three analytical pillars, risk weighting, detectability filtering, and multi-criteria coverage scoring, extend the sector-specific evaluation model of [13] while leveraging both the Enterprise and ICS ATT&CK matrices to ensure parity across IT and OT surfaces [15].

### 3.1 Phase 1: Energy Sector Threat Intelligence Collection

Phase 1 systematically collected threat intelligence by identifying attack techniques from documented cybersecurity incidents in the energy sector. Leveraging the MITRE ATT&CK framework, the analysis targeted entities with established relevance to the industry, drawing technique data directly from the official *Enterprise ATT&CK* repository [4] and *ICS ATT&CK* repository [5].

*Entities Selected for Analysis*

- Software/Malware Families:
  Stuxnet (S0603), Industroyer (S0604), TRITON (S1009), TrickBot (S0266), LockerGoga (S0372), NotPetya (S0368), and WannaCry (S0366).
- Campaign Activities:
  Ukraine Electric Power Attacks—2015 (C0028), 2016 (C0025), and 2022 (C0034).
- Threat-Actor Groups:
  Sandworm Team (G0034), Lazarus Group (G0032), Dragonfly 2.0 (G0074), OilRig (G0049), and TEMP. Veles (G0088).

Python scripts mined technique relationships, cross-matrix overlaps, removed duplicates and validated matches against incident reports, yielding 274 unique techniques.

### 3.2 Phase 2: Risk Assessment and Prioritization

Risk assessment employed a multi-dimensional approach combining empirical frequency analysis with established risk-calculation methodologies to prioritise techniques according to both prevalence and potential impact. Frequency scoring reflected technique prevalence across documented energy-sector incidents and was computed with logarithmic scaling:

$$frequency\_score = \log_2\left(campaign\_occurrences + 1\right).$$

This logarithmic transformation prevents high-frequency techniques from dominating the calculation while still giving greater weight to techniques with multiple occurrences. Resulting frequency scores ranged from 1 to 5 across the analysed technique set.

---

[4] (https://raw.githubusercontent.com/mitre/cti/master/enterprise-attack/enterprise-attack.json)
[5] (https://raw.githubusercontent.com/mitre/cti/master/ics-attack/ics-attack.json)

Base risk scores were derived using Pekka Hagström's *MITRE ATT&CK Risk Index Calculator* methodology [11], which quantifies risk in terms of

- Technique characteristics: complexity, stealth, and detection difficulty,
- Impact potential: severity of consequences if successfully executed,
- Mitigation availability: ease of implementing defensive measures.

Combining frequency scores with the base-risk calculations yielded weighted TTP risk values from 2.1 to 10.8, enabling prioritisation of techniques by both empirical prevalence and theoretical impact.

### 3.3 Phase 3: Detectability Assessment

The detectability assessment primarily focused on ascertaining whether specific attack techniques could be effectively observed and identified through network-based intrusion detection systems, thereby filtering the initial set of techniques to retain only those amenable to network-level detection. An automated classification process was implemented utilizing Python regular expressions and pandas DataFrame operations to process the MITRE ATT&CK data sources systematically. Techniques were categorized as "network-detectable" if their documented data sources included terms such as "Network," "Packet," "Protocol," "ICS Network," or "Network Traffic." Conversely, techniques were classified as "host/physical-based" if their primary data sources referenced terms like "Process," "File," "Registry," "Memory," "Sensor," "Kernel," or "Application Log." It is important to note that some techniques presented "Partial" detectability, indicating possibilities for both network and host-based detection. This rigorous detectability assessment ultimately reduced the initial compilation of 274 techniques to a refined set of 94 network-detectable techniques. This rigorous filtering process ensured that all subsequent dataset coverage analyses concentrated exclusively on techniques possessing viable network-based detection signatures, aligning precisely with the study's central emphasis on network intrusion detection systems.

### 3.4 Phase 4: Dataset Knowledge Base Construction and Characterisation

Five IDS datasets were selected for their relevance to energy sector cybersecurity and research availability. These datasets represent diverse technological domains and attack coverage. CIC-IDS2017 focuses on enterprise IT network attacks with comprehensive labeling. SWaT covers 36 cyber-physical attacks in a water treatment plant for OT, and WADI includes 14 distribution network attacks. CIC-Modbus2023 addresses industrial protocol vulnerabilities with 9 Modbus/TCP substation attack scenarios. Sherlock offers a hybrid IT/OT view with multi-stage power grid attacks, specifically covering the IEC 60870-5-104 protocol. These datasets collectively provide a broad foundation for evaluating IDS capabilities across critical energy infrastructure.

Each dataset was characterized by reviewing relevant literature and documentation, collecting information on: Domain Classification (Enterprise IT, Industrial OT, or hybrid); Attack Type Coverage (specific scenarios and threat categories); Protocol Support (enterprise vs. industrial protocols like Modbus, DNP3, IEC 61850, IEC 60870-5-104, OPC-UA); and Technical Capabilities (network flow analysis, detection functionalities, attack diversity, and IT/OT integration).

### 3.5 Phase 5: Automated IDS Dataset Coverage Mapping

Coverage mapping employs a dual large language model (LLM) assessment to evaluate dataset effectiveness against the 94 network-observable attack techniques, providing systematic and scalable coverage analysis.

*LLM Framework Design:* The framework utilizes two independent LLMs to minimize assessment bias and improve reliability. Low-temperature settings should be employed to promote deterministic responses and reduce variability across evaluation runs. Each model receives standardized inputs comprising technique descriptions and structured dataset profiles to ensure consistent evaluation criteria.

*Assessment Framework:* For each technique-dataset pair, models evaluate five coverage dimensions: (1) Attack type representation within dataset scenarios; (2) Network protocol alignment between technique requirements and dataset capabilities; (3) Operational domain relevance matching technique context; (4) Feature granularity sufficiency for effective detection; (5) Example adequacy ensuring more than token attack representations.

*Scoring Methodology:* The framework employs a quantitative scoring approach where each affirmative assessment contributes equally to a composite coverage score ranging from 0.0 to 1.0. This continuous score is then translated into ordinal classifications to facilitate interpretation and comparison across datasets.

*Validation Principles:* When LLM assessments diverge, the framework should employ conservative classification to prevent capability overstatement. This approach ensures reported coverage levels represent minimum viable detection capabilities rather than optimistic projections, maintaining assessment reliability across different evaluation contexts.

## 4 Experimental Evaluation: Case-Study Application

This section applies the five-phase coverage-gap framework to the selected datasets, including CIC-IDS2017, SWaT, WADI, Sherlock, and CIC-Modbus2023, and reports quantitative results together with a comparative analysis of coverage patterns for energy-sector attack techniques.

### 4.1 Experimental Setup

The experimental evaluation systematically assessed 94 network-detectable techniques extracted from documented energy sector incidents against five prominent

NIDS datasets: CIC-IDS2017, SWaT, WADI, CIC-Modbus2023, and Sherlock. The evaluation employed a dual-LLM validation approach to ensure reliability and minimize assessment bias. The five datasets were selected based on their relevance to energy sector cybersecurity research and public availability. CIC-IDS2017 represents enterprise IT network attacks with comprehensive labeling. SWaT provides 36 cyber-physical attacks in water treatment environments, while WADI covers 14 distribution network attacks. CIC-Modbus2023 addresses industrial protocol vulnerabilities with 9 Modbus/TCP substation scenarios. Sherlock offers hybrid IT/OT coverage with multi-stage power grid attacks using IEC 60870-5-104 protocol. For LLM Validation Configuration, Claude (Anthropic) was the primary assessment engine, with Gemini (Google) providing cross-validation. Both systems operated with temperature settings of 0.1 to ensure consistent responses. Rate limiting was implemented to maintain API stability and reproducibility.

### 4.2 Coverage Assessment Implementation

*Processing Parameters:* Techniques were processed in batches of five per API call to optimize assessment quality. Each technique underwent independent evaluation by both LLM systems to establish inter-model reliability metrics.

*Evaluation Criteria:* For every technique–dataset pair the models answer five yes/no questions: (i) does the dataset already contain a comparable attack type? (ii) does it record the protocol the technique relies on? (iii) was it captured in the same operational domain (enterprise IT or industrial OT)? (iv) does it expose the packet or process features needed to detect the attack? and (v) does it offer more than token examples of the attack? Each affirmative answer adds 0.2, yielding a coverage score between 0.0 and 1.0.

*Classification System:* This score is converted into four per-dataset labels: $FULL_cOV\ ERAGE$ (1.0), $PARTIAL_cOV\ ERAGE$ (0.5), $NO_cOV\ ERAGE$ (0.0), or UNKNOWN (0.25 when evidence is inconclusive). When Claude and Gemini disagree, the lower (more cautious) label is kept, ensuring coverage is never overstated.

## 5 Results and Discussion

### 5.1 Dataset Coverage Analysis Results

Sherlock achieved the highest mean coverage score at 0.563, with 35 techniques (37.2%) receiving complete coverage classification. CIC-IDS2017 followed closely with a mean score of 0.547 and 26 techniques (27.7%) achieving full coverage. CIC-Modbus2023 demonstrated moderate performance with a 0.516 mean score and 32 techniques (34.0%) fully covered. Both SWaT and WADI exhibited identical performance patterns, each achieving 0.378 mean coverage with 11 techniques (11.7%) fully covered.

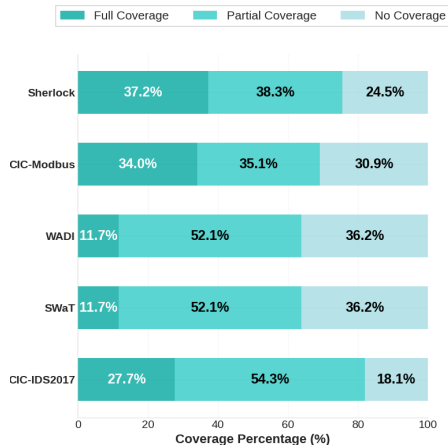

**Fig. 3.** Dataset coverage distribution

Dataset coverage results are presented in Figure 3. This reflects Sherlock's explicit modeling of IEC 60870-5-104 traffic relevant to power grids, while CIC-IDS2017 captures broad IT techniques often used as entry points in real incidents. In contrast, SWaT and WADI underperform due to their narrow, plant-specific scope.

### 5.2 Dataset Complementarity Analysis

The additive coverage potential of combining datasets was explored. Two-dataset combinations showed that CIC-IDS2017 + Sherlock offered the best combined coverage at 87%, followed by CIC-Modbus2023 + CIC-IDS2017 at 84%, and SWaT + CIC-IDS2017 at 79%. For three-dataset combinations, CIC-IDS2017 + Sherlock + CIC-Modbus2023 achieved 92% combined coverage. Using all five datasets resulted in 94% combined coverage, missing only one technique. Optimization recommendations suggest that a three-dataset combination (CIC-IDS2017 + Sherlock + CIC-Modbus2023) provides 92% coverage with optimal resource utilization, serving as a minimum viable coverage. While all five datasets are required for maximum coverage, additional datasets yield diminishing returns. This experimental evaluation demonstrates significant coverage gaps in individual datasets while emphasizing the complementary nature of different dataset types for comprehensive energy sector threat detection.

Coverage Distribution Analysis: No single technique achieved full coverage across all five datasets, indicating fundamental gaps in the collective dataset ecosystem. Partial coverage dominated the results, with 93 out of 94 techniques (98.9%) receiving at least one *PARTIAL_COV ERAGE* classification across the dataset collection. Only one technique (1.1%) experienced complete coverage gaps across all datasets.

The analysis identified several critical coverage gaps with significant energy sector cybersecurity research implications. The universal absence of coverage for T1021.002 (Remote Services: SMB/Windows Admin Shares) across all datasets represents a fundamental limitation, as this technique is crucial for lateral movement in energy infrastructure attacks. Similarly, techniques T0804 (Program PLC), T0805 (Manipulate Control), and T1570 (Lateral Tool Transfer) showed minimal coverage, despite their documented use in energy sector incidents. These gaps are particularly concerning given the documented attack patterns in energy sector incidents. The 2015 and 2016 Ukraine power grid attacks demonstrated the critical importance of lateral movement techniques and industrial control manipulation, yet current datasets inadequately represent these attack vectors. This misalignment between documented threats and available training data creates a significant risk for detection systems developed using these datasets.

### 5.3 Critical Coverage Gaps and Implications

The analysis also reveals systematic biases in dataset construction. Enterprise IT datasets like CIC-IDS2017 excel in network service exploitation but lack industrial protocol coverage, while OT-focused datasets like SWaT and WADI provide strong coverage of process-level attacks but miss enterprise pivot techniques. This specialization, while understandable from a research perspective, creates blind spots when considering the end-to-end attack chains typical of energy sector incidents. The correlation analysis is shown in Figure 4.

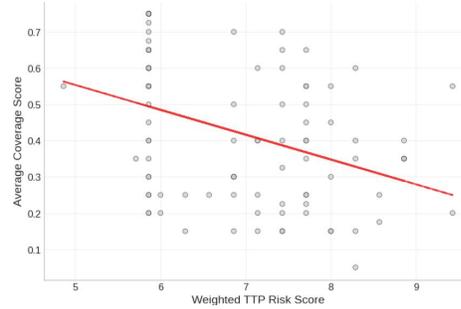

**Fig. 4.** Coverage vs. risk-score correlation

### 5.4 Inter Model Analysis

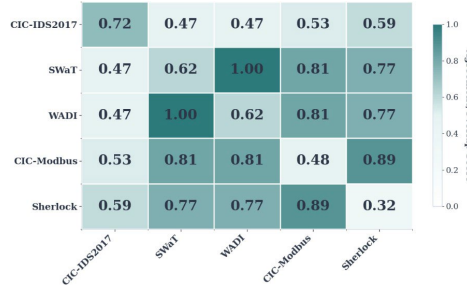

**Fig. 5.** Dataset coverage agreement matrix (diagonal: Claude–Gemini agreement, off diagonal: dataset overlap)

Analysis of agreement rates between Claude and Gemini LLMs revealed dataset-specific variations. Agreement rates were 72.3 for CIC-IDS2017, 61.7 for SWaT, 61.7 for WADI, 47.9 for CIC-Modbus2023, and 31.9 for Sherlock. The overall agreement across all dataset-technique combinations was 51.9.

High-agreement scenarios included enterprise IT techniques in CIC-IDS2017, basic network protocols across all datasets, and well-documented attack techniques. Low-agreement scenarios involved hybrid IT/OT techniques on specialized datasets, novel or emerging attack patterns, and techniques that required domain-specific knowledge. A conservative approach was adopted when the models disagreed, selecting the lower of the two coverage classifications. The LLM agreement analysis is summarized in Figure 5. High agreement for IT-centric datasets confirms stable representation, while low agreement for hybrid datasets highlights uncertainty in emerging IT/OT scenarios, underscoring the need for cautious interpretation.

# 6 Conclusion

Publicly available benchmark datasets for developing AI-based Network Intrusion Detection Systems (NIDS) are typically created with a focus on enterprise IT environments, which limits their suitability for the complex, converged IT and OT architectures prevalent in the energy sector. This study investigates the representativeness of five such datasets: CIC-IDS2017, SWaT, WADI, Sherlock, and CIC-Modbus2023, analyzing their coverage of 94 network-detectable MITRE ATT&CK techniques drawn from 274 documented cyber incidents in the energy domain. A structured evaluation framework was used to assess the protocol's diversity, the scenarios' representation, and the detection's relevance. The results show that Sherlock and CIC-IDS2017 offer the highest individual coverage (0.56 and 0.55), while SWaT and WADI trail by 0.38. However, when strategically combined, CIC-IDS2017, Sherlock, and CIC-Modbus2023 cover 92% of the relevant techniques. Key gaps persist in representing lateral movement and manipulation of industrial protocols. These findings emphasize the pressing need for the expansion of the dataset to facilitate a realistic evaluation of NIDS in energy sector contexts and provide a reproducible methodology to guide future dataset development.